\documentclass[3p,times,procedia]{elsarticle}
\usepackage{nupha_ecrc}
\usepackage{lineno}

\volume{00}
\firstpage{1}
\journalname{Nuclear Physics A}
\runauth{}
\jid{nupha}
\jnltitlelogo{Nuclear Physics A}

\usepackage{amssymb}
\usepackage[figuresright]{rotating}

\begin{document}
\begin{frontmatter}

\dochead{XXVIIth International Conference on Ultrarelativistic Nucleus-Nucleus Collisions\\ (Quark Matter 2018)}
\title{Quarkonium production in p-Pb collisions with ALICE}
\author{Biswarup Paul for the ALICE collaboration}
\address{University and INFN Torino, Italy \\ Via Pietro Giuria 1, I-10125 Torino, Italy}

\begin{abstract}
ALICE has measured quarkonium production in p-Pb collisions at backward ($-$4.46 $<$ $y_{\rm cms}$ $<$ $-$2.96), mid ($-$1.37 $<$ $y_{\rm cms}$ $<$ 0.43) and forward (2.03 $<$ $y_{\rm cms}$ $<$ 3.53) rapidity ($y$) regions down to zero transverse momentum ($p_{\rm T}$). The inclusive J/$\psi$ production has been studied at mid-$y$ in p-Pb interactions at $\sqrt{s_{\rm NN}}$ = 5.02 TeV and at forward and backward $y$ in p-Pb collisions at $\sqrt{s_{\rm NN}}$ = 5.02 TeV and 8.16 TeV. The comparison of the J/$\psi$ production to the one of the loosely bound $\psi$(2S) state is discussed, together with new results on the nuclear modification factors of the $\Upsilon$(1S) and $\Upsilon$(2S) states measured at forward and backward $y$. All the results will be compared to those obtained at lower energies and with available theoretical calculations.
\end{abstract}

\end{frontmatter}

%\linenumbers

%%%%%%%%%%%%%%%%%%%%%%%%%%%%%%%%%%%%%%%%%%%%%%%%%%%%%%
\section{Introduction}
%%%%%%%%%%%%%%%%%%%%%%%%%%%%%%%%%%%%%%%%%%%%%%%%%%%%%%
The study of quarkonium production in proton-nucleus collisions is an important tool to investigate cold nuclear matter (CNM) effects. Mechanisms such as the modification of the parton distribution functions in nuclei, the presence of a color glass condensate or coherent energy loss of the $c\overline{c}$ or $b\overline{b}$ pair in the medium have been employed to describe the results on J/$\psi$ and $\Upsilon$ production obtained in proton-nucleus collisions from the LHC Run 1~\cite{Jpsi5TeV_pt_y, Jpsi5TeV_cent, Upsi5TeV}. In addition, final state mechanisms, possibly related to the presence of a dense medium, are required to explain the stronger suppression observed for the loosely bound $\psi$(2S) state~\cite{Psip5TeV_pt_y, Psip5TeV_cent}. The measurement of the inclusive J/$\psi$ $v_{2}$ is done via a study of the angular correlations between forward and backward J/$\psi$ and mid-rapidity charged particles~\cite{Jpsi_v2_pPb}. A strong indication of long-range correlations with a sizeable non-zero $v_{2}$ at high transverse momentum is comparable to the one already observed in Pb-Pb collisions, suggesting a common mechanism. The larger statistics collected in LHC Run 2 allow us a more detailed study of the quarkonium production in p-Pb collisions, at both $\sqrt{s_{\rm NN}}$ = 5.02 and 8.16TeV, providing further insight on the involved cold nuclear matter mechanisms.

%%%%%%%%%%%%%%%%%%%%%%%%%%%%%%%%%%%%%%%%%%%%%%%%%%%%%%
\section{Experimental setup and data analysis}
%%%%%%%%%%%%%%%%%%%%%%%%%%%%%%%%%%%%%%%%%%%%%%%%%%%%%%
The ALICE Collaboration has studied inclusive quarkonium production in p-Pb collisions at mid-$y$ in the dielectron channel and at forward/backward-$y$ in the dimuon channel. Due to the beam-energy asymmetry during the p-Pb data-taking, the nucleon-nucleon center-of-mass system is shifted in rapidity with respect to the laboratory frame by $\Delta y$ = 0.465 towards the proton beam direction. The data have been taken with two beam configurations, obtained by inverting the directions of the p and Pb beams. Since muons are identified and tracked in the Muon Spectrometer, which covers the pseudorapidity range $-4<\eta<-2.5$~\cite{alice}, this results in a forward (2.03 $<$ $y_{\rm cms}$ $<$ 3.53) and backward ($-$ 4.46 $<$ $y_{\rm cms}$ $<$ $-$ 2.96) accessible rapidity regions.  Mid rapidity coverage is $-$ 1.37 $<$ $y_{\rm cms}$ $<$ 0.43. The Silicon Pixel Detector (SPD) is used for vertex identification. The V0 detector provides the minimum-bias trigger and helps to remove the beam-induced background. Two sets of Zero Degree Calorimeters (ZDCs), each including a neutron (ZN) and a proton (ZN) calorimeter, are used for the centrality estimation. The centrality selection is defined by a selected range of energy deposited by neutrons in the Pb-remnant side of ZN using the hybrid method described in~\cite{centrality}. In this method, the determination of the average number of binary nucleon collisions $\langle N_{\rm coll} \rangle$ relies on the assumption that the charged-particle multiplicity measured at mid-rapidity is proportional to the number of participant nucleons ($\langle N_{\rm part} \rangle$). $\langle N_{\rm part} \rangle$ is calculated from the Glauber model~\cite{glauber} which is generally used to calculate geometrical quantities of nuclear collisions. Other assumptions to derive $\langle N_{\rm coll} \rangle$, which are discussed in~\cite{centrality}, are used in order to determine the associated systematic uncertainty. The centrality classes 0-2\% and 90-100\% are excluded due to the possible contamination from residual pile-up events. Events where two or more interactions occur in the same colliding bunch (in-bunch pile-up) or during the readout time of the SPD (out-of-bunch pile-up) are removed using the information from SPD and V0. More details on the experimental apparatus can be found in~\cite{alice}. Details on the analysis techniques and event selection are reported in Ref.~\cite{pPbJpsi_pt_y, pPbJpsi_cent, pPbJpsi_mid, pPbJpsi_mid_paper, pPbUpsi_pt_y_cent}.

%%%%%%%%%%%%%%%%%%%%%%%%%%%%%%%%%%%%%%%%%%%%%%%%%%%%%%
\section{Results}
%%%%%%%%%%%%%%%%%%%%%%%%%%%%%%%%%%%%%%%%%%%%%%%%%%%%%%
The nuclear modification factor ($R_{\rm pPb}$) is defined as the ratio of quarkonium production yield in p-Pb collisions to that in pp collisions collected at the same center-of-mass energy scaled with number of binary nucleon-nucleon collisions. For centrality dependent studies in p-Pb collisions in ALICE it is referred to as $Q_{\rm pPb}$ due to the possible bias in the determination of centrality.

%%%%%%%%%%%%%%%%%%%%%%%%%%%%%%%%%%%%%%%%%%%%%%%%%%%%%%
%                      Jpsi and Psi(2S)
%%%%%%%%%%%%%%%%%%%%%%%%%%%%%%%%%%%%%%%%%%%%%%%%%%%%%%
\begin{figure}[ht]
\center
\includegraphics[scale=0.374]{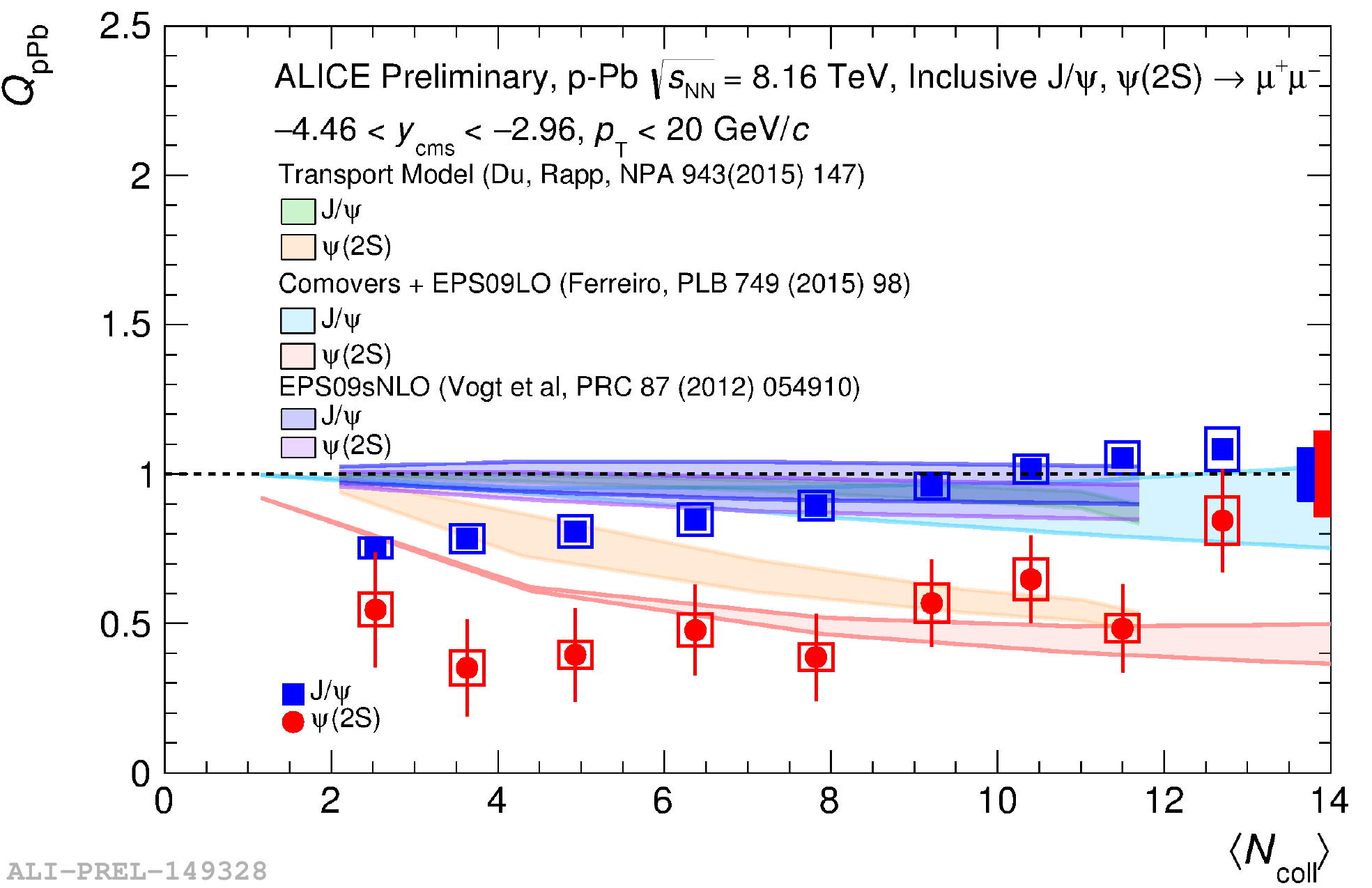}
\includegraphics[scale=0.374]{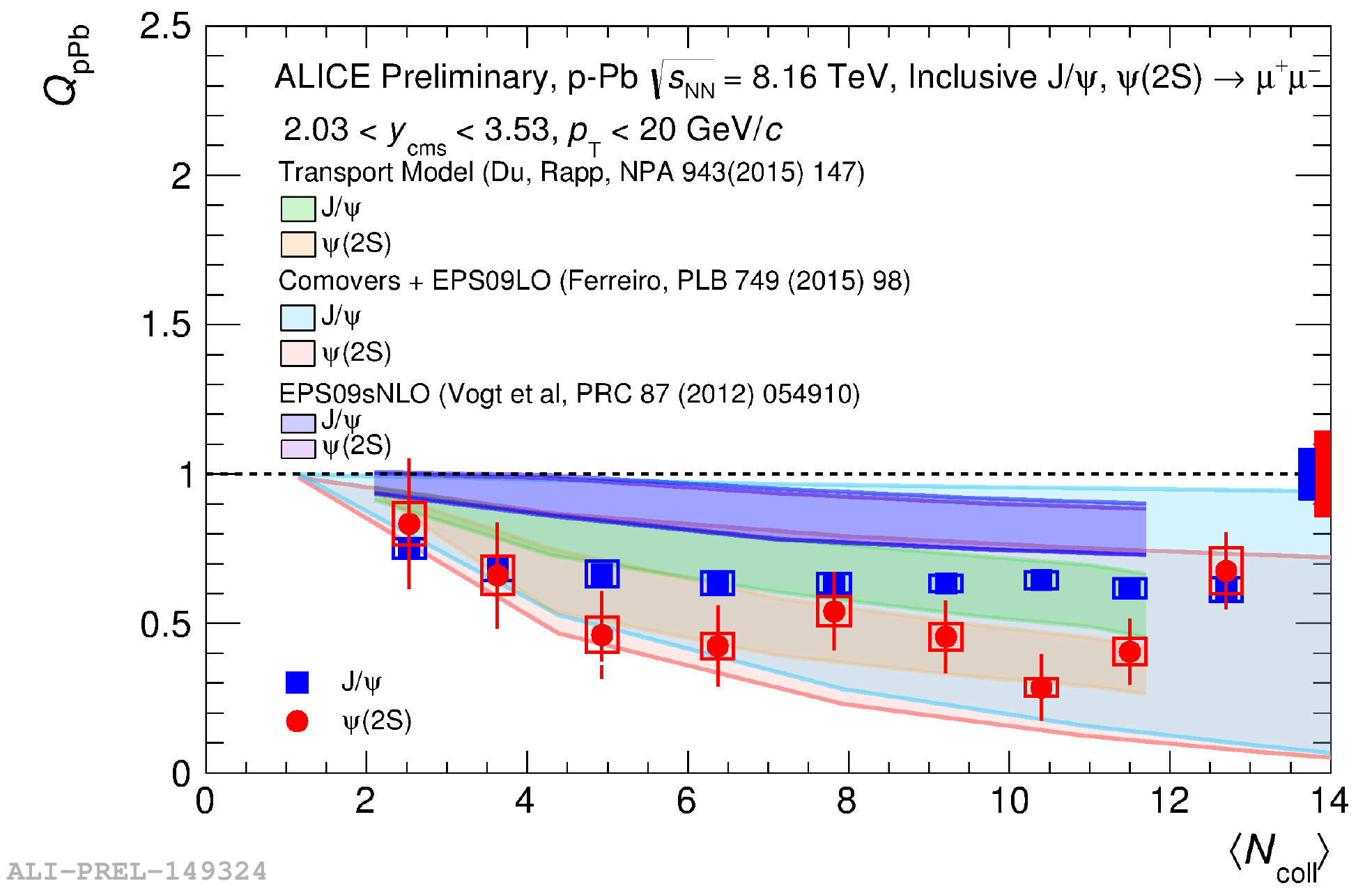}
\caption{\label{QpPb8TeV_psip}$Q_{\rm pPb}$ of inclusive J/$\psi$ and $\psi$(2S) as a function of $\langle N_{\rm coll} \rangle$ at backward (left) and forward (right) rapidity at $\sqrt{s_{\rm{NN}}}$ = 8.16 TeV.}
\end{figure}

$Q_{\rm pPb}$ of J/$\psi$ and $\psi$(2S) as a function of $\langle N_{\rm coll} \rangle$ are shown in Fig.~\ref{QpPb8TeV_psip}. The J/$\psi$ $Q_{\rm pPb}$ shows a reduction from peripheral to central collisions at forward-$y$, while trend is opposite at backward-$y$. Measurement at mid-$y$, reported in~\cite{pPbJpsi_mid}, shows almost no centrality dependence of $Q_{\rm pPb}$. $\psi$(2S) suppression is stronger than J/$\psi$ especially at backward-$y$. The results are compared to a pure nuclear shadowing theory calculation~\cite{eps09} based on EPS09s NLO set of nuclear parton distribution functions (nPDFs). The model is in fair agreement with both J/$\psi$ and $\psi$(2S) results at forward-$y$ but can not describe the result at backward-$y$. At backward-$y$, final state effects are needed to explain the $\psi$(2S) behaviour~\cite{comovers,transport}. Theoretical predictions based on a comover approch with EPS09LO set of nPDF~\cite{comovers} and on a transport model~\cite{transport}, which includes CNM effects and the interaction with the produced medium describe the backward-$y$ results quite well although some discrepancies are observed between the data and the models in the peripheral collisions.

%%%%%%%%%%%%%%%%%%%%%%%%%%%%%%%%%%%%%%%%%%%%%%%%%%%%%%
%                     Jpsi mult
%%%%%%%%%%%%%%%%%%%%%%%%%%%%%%%%%%%%%%%%%%%%%%%%%%%%%%
\begin{figure}[ht]
\center
\includegraphics[scale=0.374]{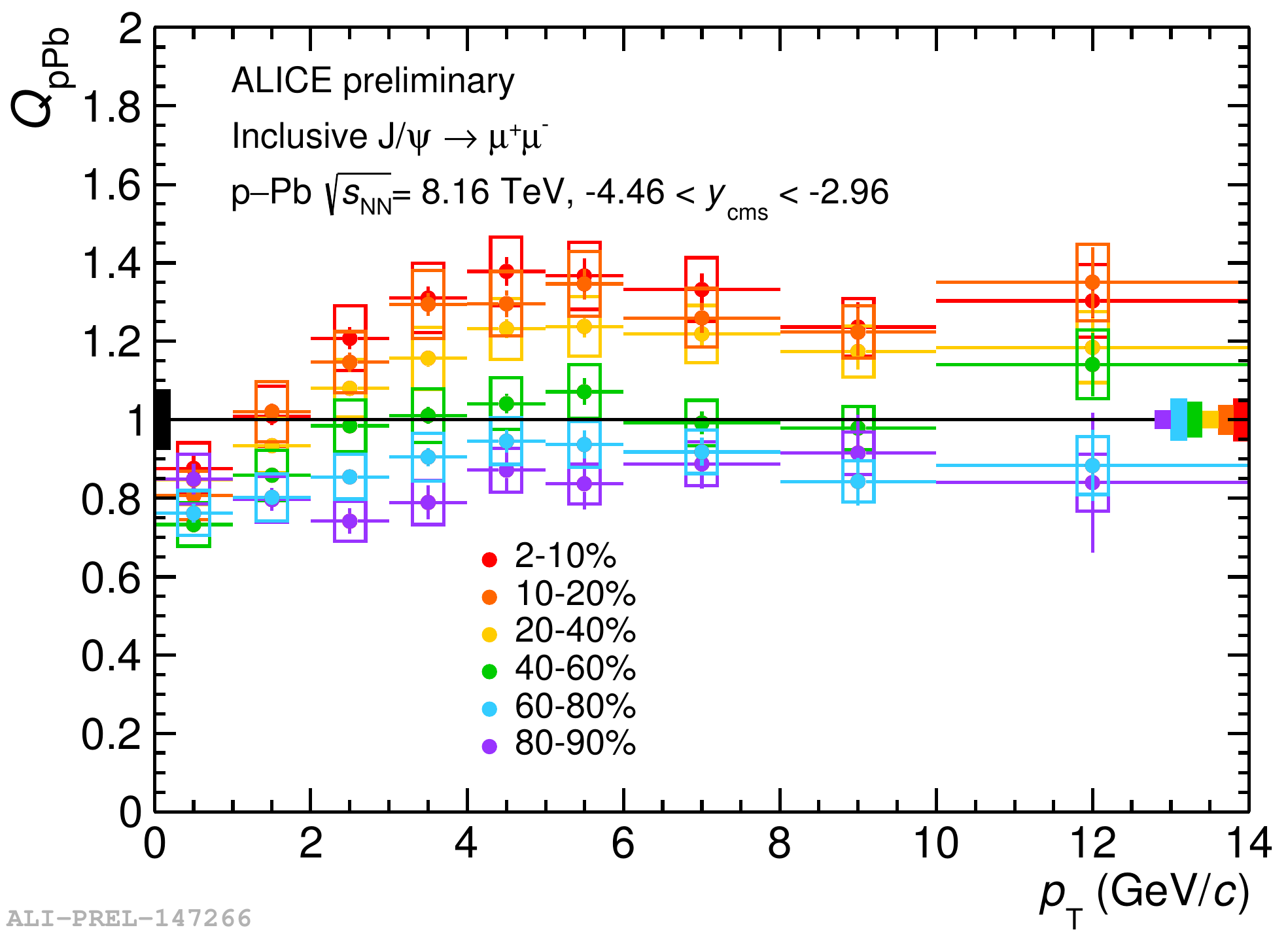}
\includegraphics[scale=0.374]{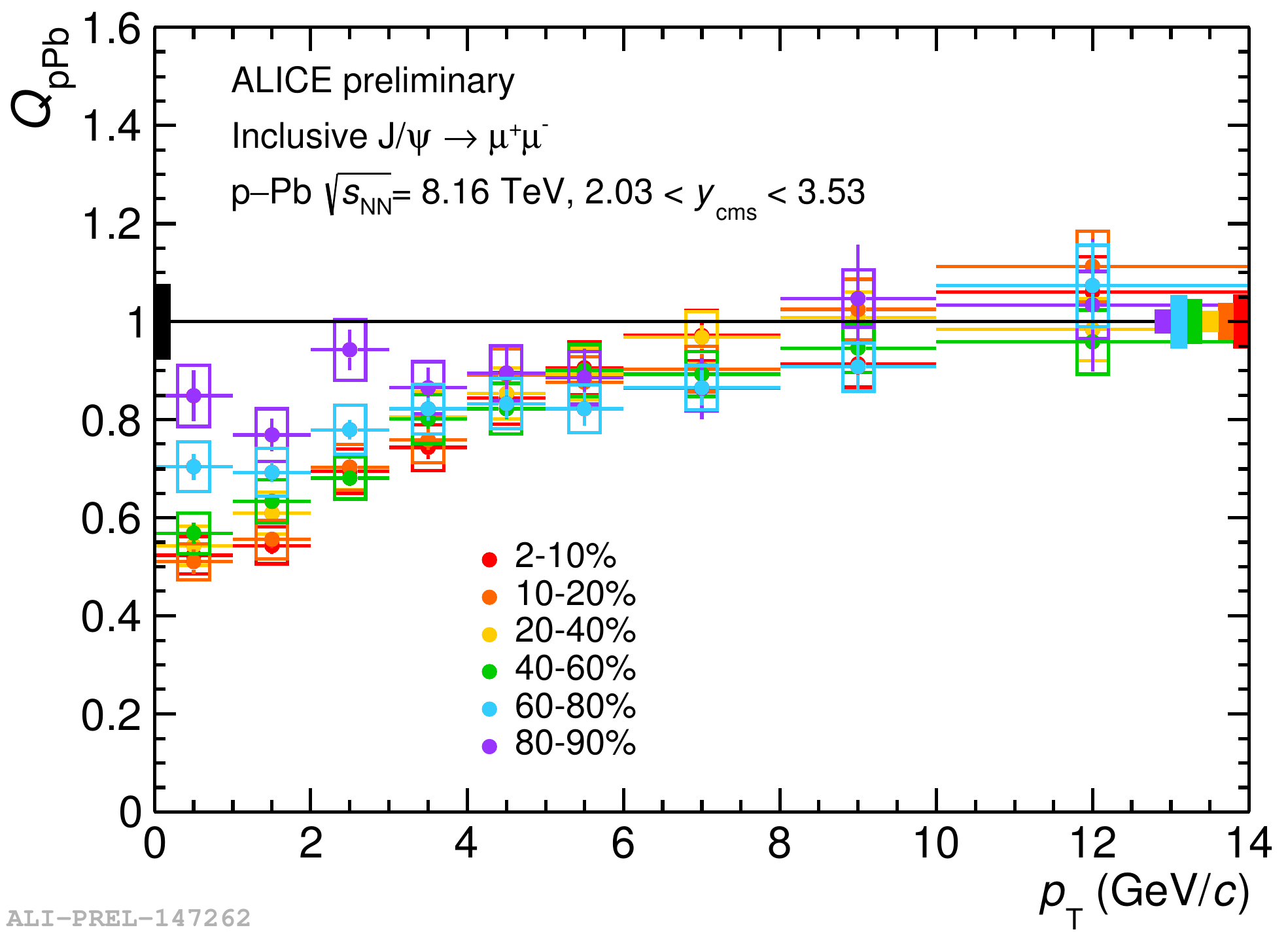}
\caption{\label{QpPb8TeV_pt}Inclusive J/$\psi$ $Q_{\rm pPb}$ as a function of $p_{\rm T}$ at backward (left) and forward (right) rapidity at $\sqrt{s_{\rm{NN}}}$ = 8.16 TeV for different centrality classes.}
\end{figure}

Fig.~\ref{QpPb8TeV_pt} shows results on multi-differential study of J/$\psi$ $Q_{\rm pPb}$ as a function of $p_{\rm T}$ in different centrality classes. A clear evolution of $Q_{\rm pPb}$ as a function of $p_{\rm T}$ in different centrality classes is observed. At backward-$y$ there is an enhancement in most central collisions for $p_{\rm T}$ $>$ 3 GeV/$c$. At forward-$y$ stronger suppression at low $p_{\rm T}$ in most central collisions is observed and $Q_{\rm pPb}$ is compatible with unity for $p_{\rm T}$ $>$ 7 GeV/$c$ within uncertainties for all centrality intervals.

%%%%%%%%%%%%%%%%%%%%%%%%%%%%%%%%%%%%%%%%%%%%%%%%%%%%%%
%                     Upsilon
%%%%%%%%%%%%%%%%%%%%%%%%%%%%%%%%%%%%%%%%%%%%%%%%%%%%%%
\begin{figure}[ht]
\center
\includegraphics[scale=0.374]{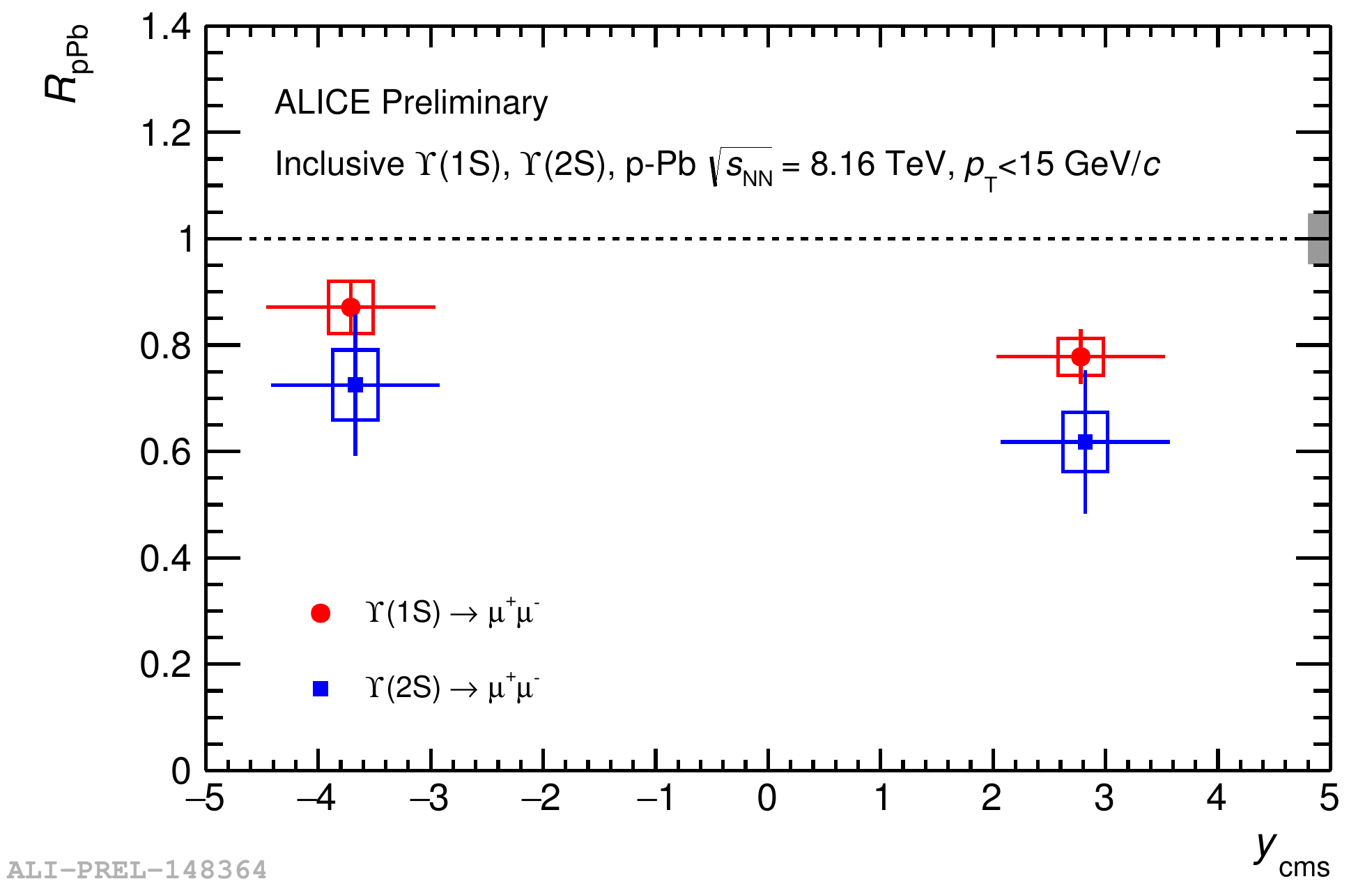}
\includegraphics[scale=0.374]{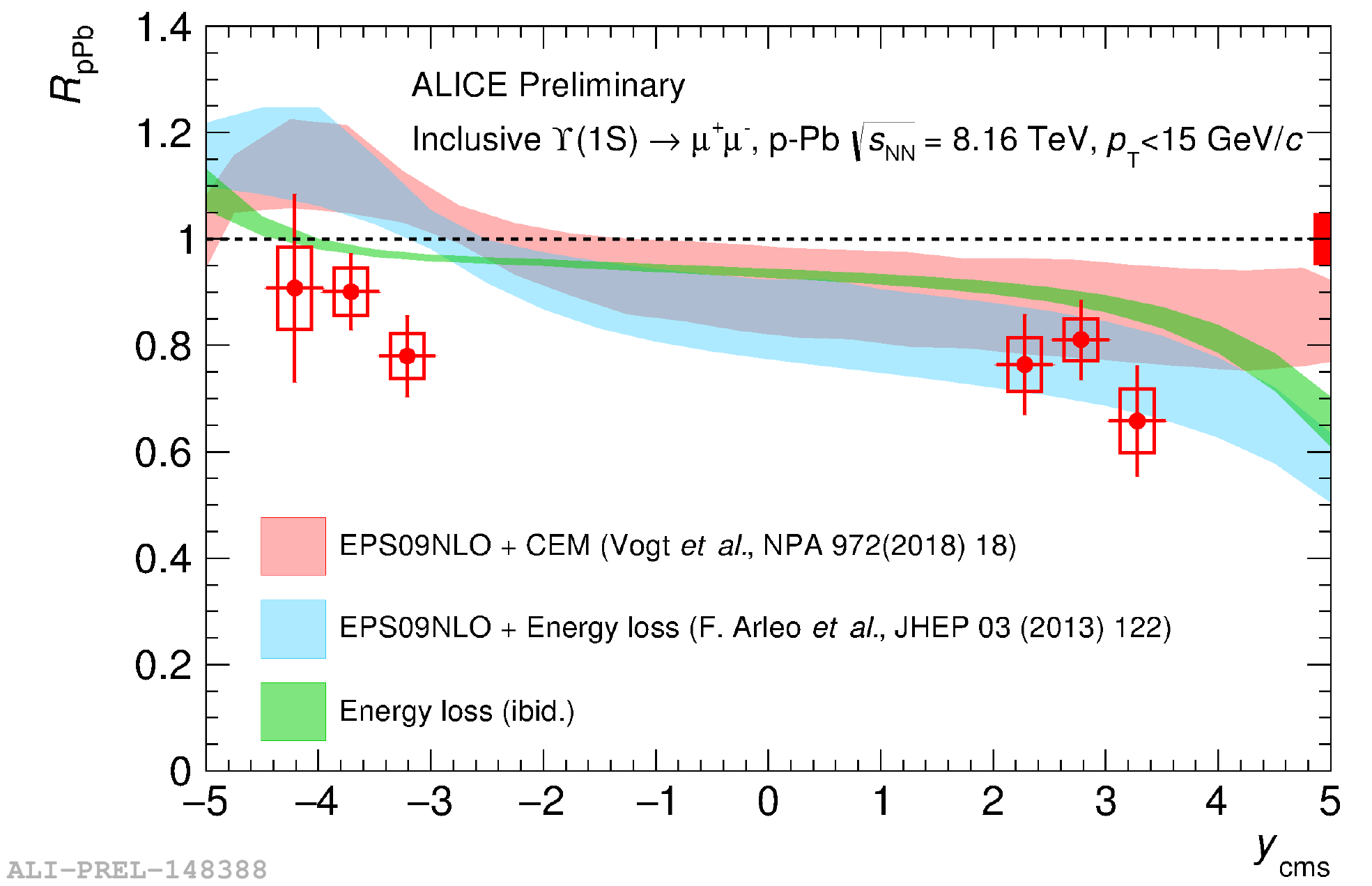}
\caption{\label{RpPb8TeV_upsi}$R_{\rm pPb}$ of inclusive $\Upsilon$(1S) and $\Upsilon$(2S) as a function of center-of-mass rapidity at $\sqrt{s_{\rm{NN}}}$ = 8.16 TeV (left). $\Upsilon$(1S) results are compared to the theoretical calculations (right).}
\end{figure}

The large statistics collected at $\sqrt{s_{\rm{NN}}}$ = 8.16 TeV allows us to measure $\Upsilon$(1S) production in rapidity, $p_{\rm T}$ and centrality bins whereas at $\sqrt{s_{\rm{NN}}}$ = 5.02 TeV~\cite{Upsi5TeV} we have results only as a function of rapidity due to low statistics. Results are compatible between the two center-of-mass energies. From Fig.~\ref{RpPb8TeV_upsi} (left) one can see that there is a suppression of the $\Upsilon$(1S) production in p-Pb collisions, both at forward-$y$ and backward-$y$, with a hint for a stronger suppression at forward-$y$. The suppression amounts to 2.8$\sigma$ and 1.7$\sigma$ at forward-$y$ and backward-$y$, respectively. $R_{\rm pPb}$ of $\Upsilon$(2S) is also shown in Fig.~\ref{RpPb8TeV_upsi} (left). The difference in the $R_{\rm pPb}$ of $\Upsilon$(2S) and $\Upsilon$(1S) amounts to 1$\sigma$ at forward-$y$ and 0.9$\sigma$ at backward-$y$. CMS~\cite{Upsi5TeV_CMS} and ATLAS~\cite{Upsi5TeV_ATLAS} measurements at mid-$y$ also show that $\Upsilon$(2S) suppression is stronger than $\Upsilon$(1S). Theoretical predictions based on shadowing~\cite{eps09} and energy loss (with or without the contribution of the EPS09 nuclear shadowing)~\cite{eloss} describe forward-$y$ $\Upsilon$(1S) results but slightly overestimate backward-$y$ results, as visible in Fig.~\ref{RpPb8TeV_upsi} (right). 

\begin{figure}[h!]
\begin{center}
\includegraphics[width=7.4cm,height=5.1cm]{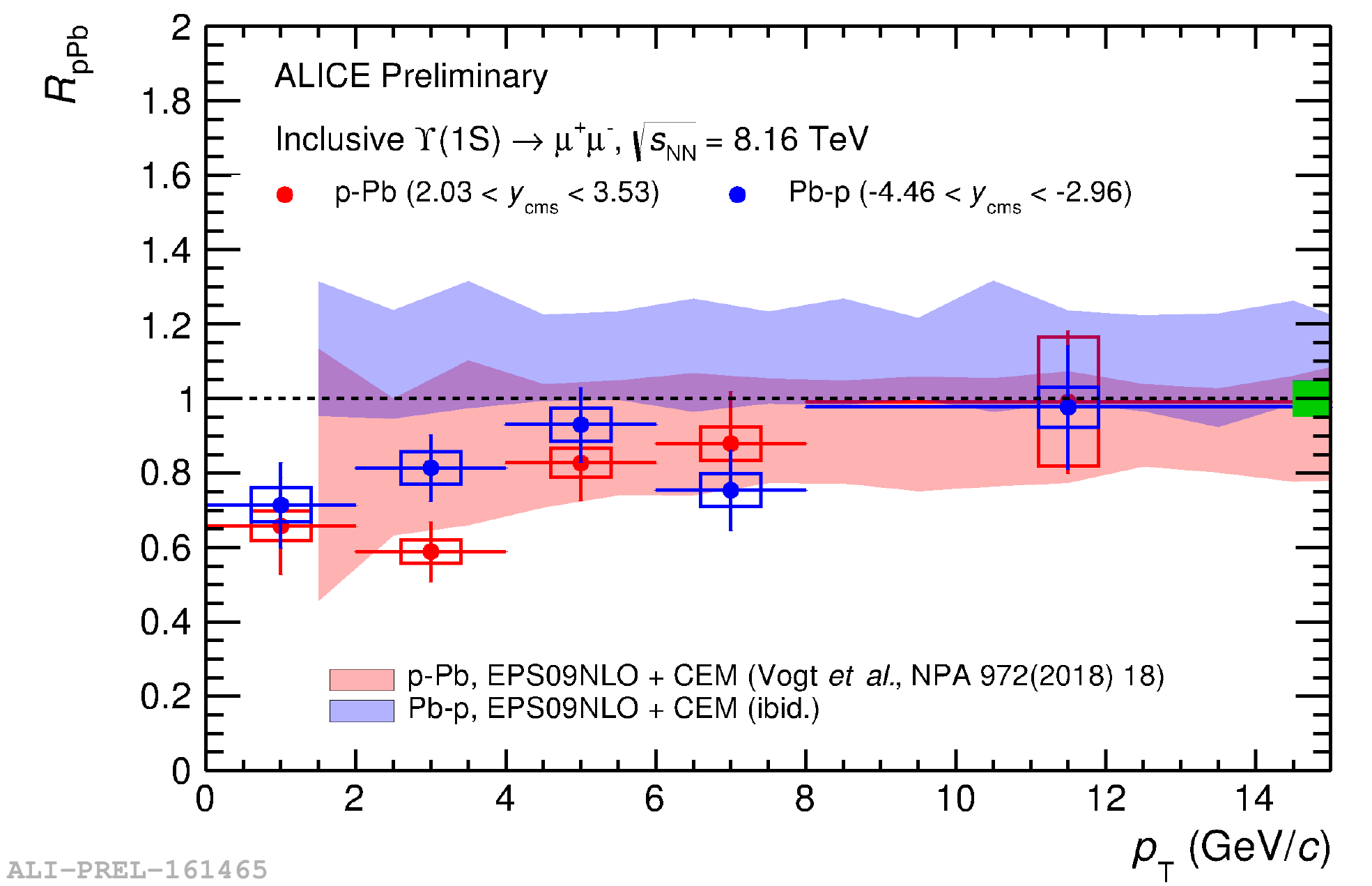}
\includegraphics[width=7.4cm,height=5.1cm]{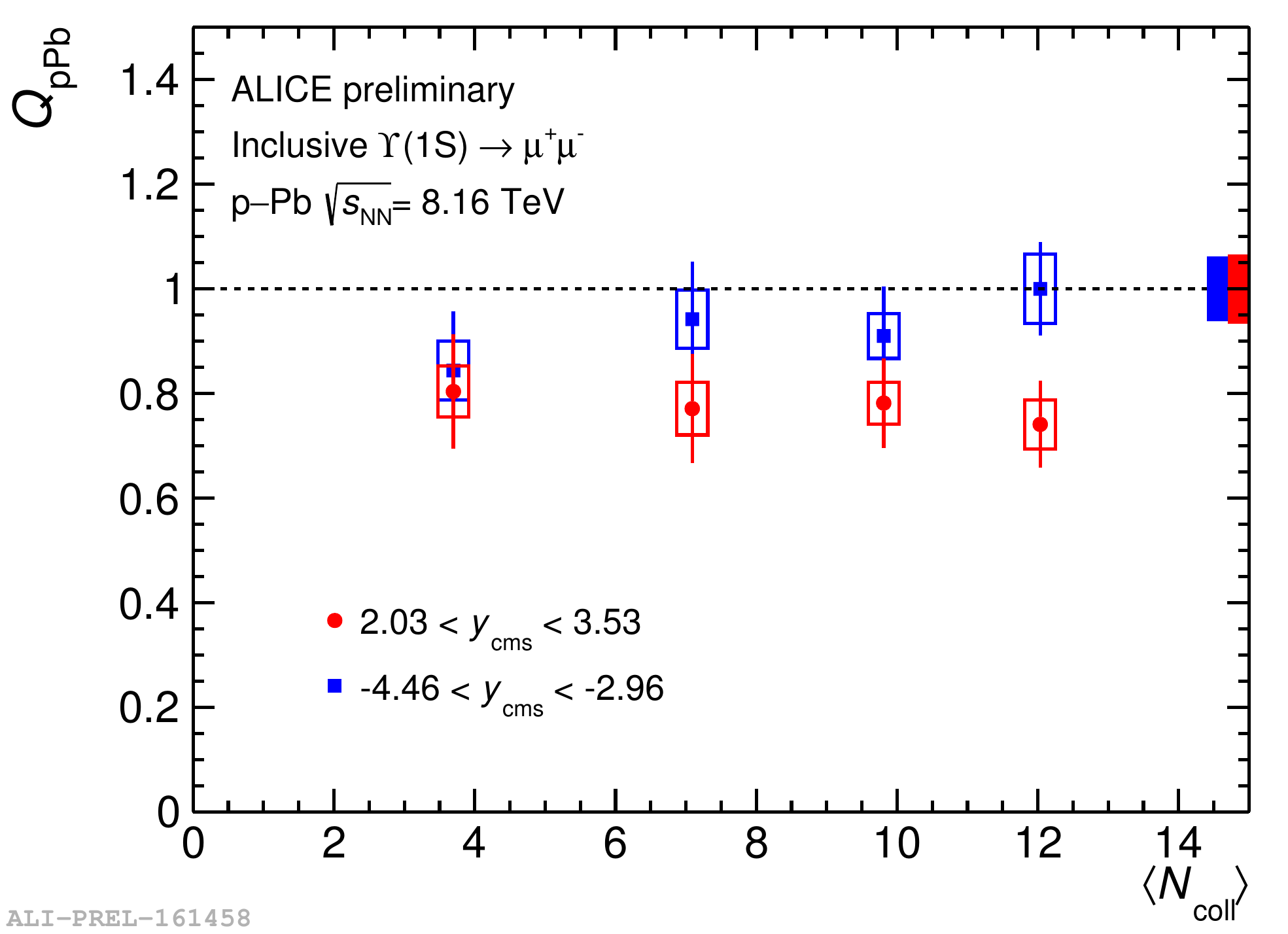}
\caption{Left: inclusive $\Upsilon$(1S) $R_{\rm pPb}$ as a function of $p_{\rm T}$ compared to the theoretical calculations. Right: inclusive $\Upsilon$(1S) $Q_{\rm pPb}$ as a function of $\langle N_{\rm coll} \rangle$ at backward and forward rapidity at $\sqrt{s_{\rm{NN}}}$ = 8.16 TeV.}
\label{RpPb8TeV_upsi_pt}
\end{center}
\end{figure}

Fig.~\ref{RpPb8TeV_upsi_pt} shows $\Upsilon$(1S) $R_{\rm pPb}$ (left) and $Q_{\rm pPb}$ (right) as a function of $p_{\rm T}$ and $\langle N_{\rm coll} \rangle$ at $\sqrt{s_{\rm{NN}}}$ = 8.16 TeV, respectively. The $R_{\rm pPb}$ shows a similar behaviour at both forward and backward-$y$ as a function of $p_{\rm T}$, with a hint for a stronger suppression at low $p_{\rm T}$. Also in this case, theoretical predictions based on shadowing~\cite{eps09} describe forward-$y$ results but slightly overestimate backward-$y$ results, where anti-shadowing is predicted to play an important role. There is almost no centrality dependence of $Q_{\rm pPb}$ both at forward and backward $y$ with a hint for a stronger suppression at forward-$y$.

%%%%%%%%%%%%%%%%%%%%%%%%%%%%%%%%%%%%%%%%%%%%%%%%%%%%%%
\section{Conclusions}
%%%%%%%%%%%%%%%%%%%%%%%%%%%%%%%%%%%%%%%%%%%%%%%%%%%%%%
Quarkonium production has been measured with ALICE in p-Pb collisions at $\sqrt{s_{\rm{NN}}}$ = 5.02 and 8.16 TeV. Run2 results significantly increase the precision of the measurements, but theoretical models still face some difficulties in describing consistently all results. J/$\psi$ shows a stronger suppression at forward-$y$ than at backward-$y$ and theoretical models based on CNM effects qualitatively describe J/$\psi$ results. $\psi$(2S) shows a stronger suppression than J/$\psi$ and final state effects are needed to explain its behaviour. New results on the $\Upsilon$ production show a similar suppression for the $\Upsilon$(1S) and the $\Upsilon$(2S) which can be described, at forward-$y$, by shadowing and energy loss models. However, these calculations tend to overestimate $\Upsilon$ yields at backward-$y$. 

\bibliographystyle{elsarticle-num}
\bibliography{<your-bib-database>}

\end{document}